\documentclass[preprint,pre,aps]{revtex4-1}
\usepackage{graphics}
\usepackage{amsmath,amssymb}
\usepackage[dvips]{graphicx}
\begin{document}

\title{  Gas-liquid phase coexistence and  crossover behavior of binary ionic
fluids with screened Coulomb interactions}

\author{O. Patsahan}

\affiliation{Institute for Condensed Matter Physics of the National
Academy of Sciences of Ukraine, 1 Svientsitskii St., 79011 Lviv,
Ukraine}

\date{\today}

\begin{abstract}
We study  the effects of an interaction range on the gas-liquid phase diagram and the crossover behavior of
a simple model of ionic fluids: an equimolar binary mixture  of
equisized hard spheres interacting  through   screened Coulomb
potentials which are repulsive between particles of the same species
and attractive between particles of different species.
Using the collective variables theory, we find explicit expressions for the relevant
coefficients of the effective  $\varphi^{4}$ Ginzburg-Landau
Hamiltonian in a one-loop approximation. Within the framework of
this approximation, we calculate the critical parameters and
gas-liquid phase diagrams for  varying
inverse screening length $z$. Both the  critical temperature scaled by the Yukawa potential contact value
and the  critical
density rapidly decrease  with an increase of  the
interaction range (a decrease of $z$) and then for $z<0.05$ they slowly approach the
values found for a restricted primitive
model (RPM).
We find that   gas-liquid coexistence
region reduces with an increase of $z$ and completely  vanishes at
$z\simeq 2.78$.
Our results
clearly show that an increase in the interaction range
leads to a
decrease of the crossover temperature. For  $z\simeq 0.01$, the
crossover temperature is the same as for the RPM.
\end{abstract}
\pacs{05.70.Fh, 64.60.De, 64.60.F-}

\maketitle
\section{Introduction}

The nature of phase separation  and criticality in ionic fluids with
the dominant Coulomb interactions (e.g., molten salts and
electrolytes in solvents of low dielectric constant) has been an
outstanding experimental and theoretical issue for many years.
Electrostatic correlations  are also known to play an important role
in  many other technologically relevant systems such as
charge-colloidal suspensions, room-temperature ionic liquids and
micellar solutions of ionic surfactants.
Now, a generally accepted idea is that the gas-liquid and
liquid-liquid critical points in ionic fluids  belong to the
universal class of a three-dimensional Ising model
\cite{Gutkowskii-Anisimov,Sengers_Shanks:09,Schroer:12}.
Nevertheless,  the crossover from the mean-field-like behavior to
the Ising model criticality when approaching the critical point
remains a  challenging problem for theory, simulations and
experiments \cite{Sengers_Shanks:09,Schroer:12}.

The most commonly studied theoretical model of ionic fluids is a restricted
primitive model (RPM), which consists  of equal numbers of equisized positively and negatively charged
hard spheres   immersed in a structureless dielectric continuum. The
RPM  undergoes a gas-liquid-like phase transition at low
temperature and low density
\cite{stell1,levin-fisher,Cai-Mol1,patsahan_ion}.
Theoretical
\cite{patsahan:04:1,ciach:06:1,Parola-Reatto:11} and
numerical
\cite{caillol-levesque-weis:02,Hynnien-Panagiotopoulos,luijten,kim-fisher-panagiotopoulos:05}
investigations of the  gas-liquid criticality  in the RPM have provided
strong evidence for an Ising universal class. However,
an issue of the width of the  critical region was not addressed in these works.
On the other hand,  the Ginzburg criterion
\cite{levanyuk,ginzburg,Chaikin_Lubensky} was used in
Ref.~\cite{Fisher_Levin:93,evans,fisher3,schroer_1,schroer} in order
to study the crossover from the mean-field to  asymptotic
regime, but
the obtained results failed to   give a clear answer to the question
of the extent of the crossover region in the model.

Recently, using the method of collective variables (CVs)
\cite{zubar,jukh,Yuk-Hol,Pat-Mryg-CM}, we have derived the
Landau-Ginzburg (LG) Hamiltonian for the model of ionic fluids which
includes, besides   Coulomb interactions, short-range  attractive
interactions \cite{Patsahan:13}. An important feature of the
developed approach  is that it enables  us to obtain  all the
relevant coefficients, including the square-gradient term,  within
the framework  of the same approximation. The Ginzburg temperature
for the RPM, calculated using this theory turned out  to be about
$20$ times smaller than for a one-component nonionic model.
Furthermore, the results obtained for the RPM supplemented by
short-range attractive interactions have  shown that the Ginzburg
temperature  approaches the value found for the RPM when the
strength of  Coulomb interactions becomes sufficiently large. These
results   suggest the key role of  Coulomb interactions in the
reduction of the crossover region. Nevertheless, the study of the
effect of an interaction range on the Ginzburg temperature is needed
in order to gain a better understanding of the crossover behavior in
ionic fluids.

In the present work, we extend the theory  to the binary  ionic model  with screened Coulomb interactions.
Specifically, we consider a two-component system of particles
labeled $1$ and $2$, such that the interaction potential between a
particle of species $\alpha$ and one of the species $\beta$ at a
distance  $r$ apart is as follows:
 \begin{eqnarray}
u_{\alpha\beta}(r) = \left\{
                     \begin{array}{ll}
                     \infty, & r<\sigma\\
                     (-1)^{\alpha+\beta}K\displaystyle\frac{\exp(-z(r/\sigma-1))}{r/\sigma},&
                     r\geqslant \sigma
                     \end{array}
              \right. \,,
\label{int-YRPM}
\end{eqnarray}
where $\alpha,\beta=(1,2)$.   For $K>0$, Eq.~(\ref{int-YRPM})
describes a symmetrical mixture of hard spheres of the same diameter
$\sigma$ in which the like particles interact through a repulsive
Yukawa potential for $r>\sigma$, and the unlike particles interact
through the opposite attractive Yukawa potential for $r>\sigma$. We
restrict our consideration  to the case  where the number densities
of species $1$ and $2$ are the same, i.e.,
$\rho_{1}=\rho_{2}=\rho/2$. For $K=(q)^{2}/\epsilon$,   the model
(\ref{int-YRPM}) is called a Yukawa restricted primitive model
(YRPM).  In this case, $q_{+}=-q_{-}=q$  is the charge magnitude and
$\epsilon$ is the  dielectric constant of the medium. In the limit
$z\rightarrow \infty$, the YRPM reduces to a hard sphere model
whereas  the RPM is recovered by taking the limit $z\rightarrow 0$.
Thus, the YRPM can provide a basis for the study of  the nature of
phase and critical behavior in ionic fluids and in partially ionic
fluids.

It is worth noting that the YRPM  is often used to model a system of oppositely
charged colloids \cite{Leunissen,Hynninen-06,Fortini:06,Bier-10}. The effective
(screened) colloid-colloid interactions in such a system  are due to
the presence of coions and counterions in the solvent. In this case,
$K$ and $z$ take the form:
$K/k_{B}T=Z^{2}\lambda_{B}/(1+\kappa_{D}\sigma/2)^{2}/\sigma$ and
$z=\kappa_{D}\sigma$,  where
$\kappa_{D}=\sqrt{8\pi\lambda_{B}\rho_{s}}$ is the inverse Debye
screening length,  $\lambda_{B}=e^{2}/\epsilon_{s}k_{B}T$ is the
Bjerrum length, $\rho_{s}$   is the salt concentration and
$\epsilon_{s}$ is the dielectric constant of the solvent. In a
colloid system,  the range of interaction can be  modified  by
changing the salt  concentration.

Whereas  the effect of an interaction range on the gas-liquid phase
separation of  a simple one-component  fluid has been  extensively studied (see
Ref.~\cite{Mendoub} and references herein), as far as we know  there are only a few
works addressing this issue for the case of the YRPM
\cite{Parola-Reatto:11,Fortini:06,Carvalho_Evans:97,Mier-Y-Teran}.
In particular, the evolution of the gas-liquid phase diagram of the YRPM as a
function of the interaction range was  theoretically studied using
the integral equation methods \cite{Carvalho_Evans:97,Mier-Y-Teran}
and the hierarchical reference theory (HRT) \cite{Parola-Reatto:11}.
The results obtained from the generalized mean-spherical
approximation (GMSA) show that both the critical density and the
critical temperature increase above the corresponding values for the
RPM when $z$ increases \cite{Carvalho_Evans:97}. Moreover, the GMSA
predicts a nonmonotonous behavior of the critical temperature as a
function of $z$: the critical temperature attains a maximum at
$z\approx 4$. In Ref.~\cite{Carvalho_Evans:97}, the attention was
focused on several values of $z$,  $z=0$, $1.5075$, $3$, $4$, $5$
and $6$,  and the gas-liquid coexistence was found for all the listed
values. The highest value for which the gas-liquid coexistence was
found within the framework of  integral equation methods is
$z=25$ with the MSA \cite{Mier-Y-Teran}. In
Ref.~\cite{Parola-Reatto:11} the main emphasis is made on the
critical behavior of the model.

Simulations predict a rich phase diagram involving a gas-liquid
phase separation as well as several crystalline phases, which is in
agreement with experimental confocal microscopy data for
charge-stabilized colloidal suspensions
\cite{Leunissen,Hynninen-06,Fortini:06,Bier-10}. These studies
indicate a sensitivity of the phase diagram of the YRPM to the variation of $z$.
Unlike theoretical predictions \cite{Carvalho_Evans:97,Mier-Y-Teran}, it is found  \cite{Fortini:06} that the gas-liquid
separation is not stable with respect to gas-solid coexistence for
$z>4$.

The  purpose of the present paper is to study the effects of the
interaction range on the gas-liquid phase diagram and the Ginzburg
temperature of the YRPM. To this end, following
Ref.~\cite{Patsahan:13}, we find analytical expressions for  all the relevant coefficients of the LG
Hamiltonian in a one-loop approximation. Based on these expressions,
first we calculate  the gas-liquid critical parameters, spinodals
and coexistence curves of the model for  $0.001\leq z\leq 2.781$. Remarkably,   there is no  gas-liquid
critical point for $z\geq 2.782$ in the approximation considered. Our
discussion also involves an analysis of the dependence of the
coefficients of the effective Hamiltonian  on  the interaction
range. Applying the Ginzburg criterion, we find that the
reduced Ginzburg temperature  decreases with an increase  of the
interaction range approaching the RPM value for $z\simeq 0.01$. The present analysis also
indicates  the presence of a tricritical point at $z=2.781$.

The paper is organized as follows.  A brief description of the
formalism is given in Sec.~2. The results for the gas-liquid phase
diagram and the critical parameters are presented in Sec.~3.  In
Sec~4 we discuss the effect of the interaction range on the crossover
behavior of the YRPM. Concluding remarks are made  in Sec.~5.

\section{Theory}

 \subsection{Functional representation of the grand partition function}

 We start with the YRPM and  present the interaction potential (\ref{int-YRPM}) in the form:
\begin{equation}
 u_{\alpha\beta}(r)=\phi^{\mathrm{HS}}(r)+\phi_{\alpha\beta}^{Y}(r),
\label{2.1}
\end{equation}
where $\phi^{\mathrm{HS}}(r)$ is the interaction potential between
the two   hard spheres of diameter $\sigma$. Thermodynamic and structural properties of the  system interacting
through  the potential $\phi^{\mathrm{HS}}(r)$ are assumed to be
known. Therefore, the one-component hard-sphere model is regarded as
the reference system. $\phi_{\alpha\beta}^{Y}(r)$ are the screened Coulomb potentials. Figure~1 shows the shape
of the interaction potentials
 $\phi_{\alpha\beta}^{Y}(r)/K$  for different values of the inverse screening length.
 \begin{figure}[h]
 \centering
 \includegraphics[height=6cm]{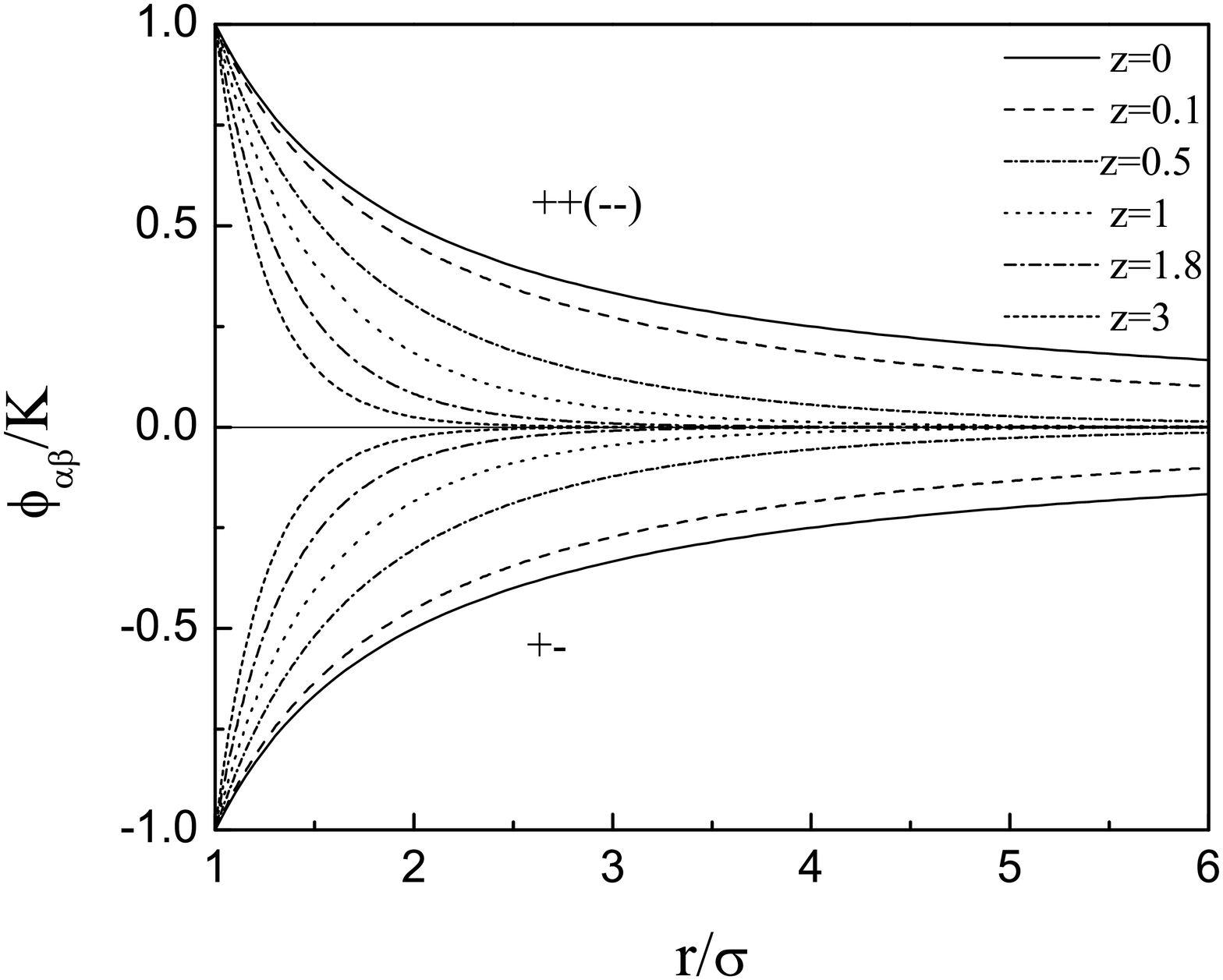}
\caption{Interaction potentials $\phi_{\alpha\beta}^{Y}(r)/K$ for different values of the inverse screening length $z$.
 } \label{fig1}
 \end{figure}

The  model under consideration is  at equilibrium in the grand
canonical ensemble, $\beta=(k_{B}T)^{-1}$ is the inverse
temperature, and $\nu_{\alpha}=\beta\mu_{\alpha}$ ($\nu_{\alpha}=\nu_{\beta}=\nu$) is the dimensionless chemical potential of the
$\alpha$th species.  Using the CV method we  present the
grand partition function  of the model  in the form of a
functional integral \cite{Patsahan:13,Pat-Mryg-CM}:
\begin{eqnarray}
 &&
\Xi=\Xi_{\text{HS}}\exp\left[\Delta\nu_{N}\langle
N\rangle_{\text{HS}}\right]
 \int ({\rm d}\rho)({\rm d}\omega)\exp\left[\Delta\nu_{N}\rho_{0,N}
-\frac{\beta}{2V}\sum_{{\mathbf k}}\widetilde \phi^{Y}(k)\rho_{{\mathbf k},Q}\rho_{-{\mathbf k},Q}
\right. \nonumber\\
&& \left. +{\rm
i}\sum_{{\mathbf k}}\left(\omega_{{\mathbf k},N}\rho_{{\mathbf
k},N}+\omega_{{\mathbf k},Q}\rho_{{\mathbf
k},Q}\right)+\sum_{n\geq 2}\frac{(-{\rm i})^{n}}{n!}\sum_{i_{n}\geq 0}^{n}
\sum_{{\mathbf{k}}_{1},\ldots,{\mathbf{k}}_{n}}
{\mathfrak{M}}_{n}^{(i_{n})}(k_{1},\ldots,k_{n})
\right. \nonumber\\
&& \left.
 \times
\omega_{{\bf{k}}_{1},Q}\ldots\omega_{{\bf{k}}_{i_{n}},Q}\,\omega_{{\bf{k}}_{i_{n+1}},N}\ldots\omega_{{\bf{k}}_{n},N}
\delta_{{\bf{k}}_{1}+\ldots +{\bf{k}}_{n}}
 \right].
 \label{Xi_full_1}
\end{eqnarray}
Here,   $\rho_{{\mathbf k},N}$
and $\rho_{{\mathbf k},Q}$ are the CVs which describe fluctuations
of the total number density and the charge density  (or relative number density), respectively:
\begin{equation*}
\rho_{{\mathbf k},N}=\rho_{{\mathbf k},+}+\rho_{{\mathbf k},-},
\qquad \rho_{{\mathbf k},Q}=\rho_{{\mathbf k},+}-\rho_{{\mathbf
k},-}.
\end{equation*}
CV $\rho_{{\mathbf k},\alpha}=\rho_{{\mathbf k},\alpha}^c-{\rm
i}\rho_{{\mathbf k},\alpha}^s$ describes  the value of the $\mathbf
k$-th fluctuation mode of the number density of the $\alpha$th
species, the indices $c$ and $s$ denote real and imaginary parts of
$\rho_{{\mathbf k},\alpha}$; CVs $\omega_{N}$ and $\omega_{Q}$ are
conjugate to  $\rho_{N}$ and $\rho_{Q}$, respectively. $({\rm
d}\rho)$ and $({\rm d}\omega)$ denote  volume elements of the CV
phase space:
\begin{displaymath}
({\rm d}\rho)=\prod_{A=N,Q}{\rm d}\rho_{0,A}{\prod_{\mathbf
k\not=0}}' {\rm d}\rho_{\mathbf k,A}^{c}{\rm d}\rho_{\mathbf
k,A}^{s}, \quad ({\rm d}\omega)=\prod_{A=N,Q}{\rm
d}\omega_{0,A}{\prod_{\mathbf k\not=0}}' {\rm d}\omega_{\mathbf
k,A}^{c}{\rm d}\omega_{\mathbf k,A}^{s}
\end{displaymath}
and the product over ${\mathbf k}$ is performed in the upper
semi-space ($\rho_{-\mathbf k,A}=\rho_{\mathbf k,A}^{*}$,
$\omega_{-\mathbf k,A}=\omega_{\mathbf k,A}^{*}$).

$\widetilde\phi^{Y}(k)$ is the Fourier transform of
the repulsive potential $\phi_{\alpha\alpha}^{Y}(r)=\phi^{Y}(r)$, where $\phi^{Y}(r)=K\sigma\exp[-z(r/\sigma-1)]/r$.
Here we use the  Weeks-Chandler-Andersen  regularization of the   potential $\phi^{Y}(r)$ inside the hard core \cite{wcha}.
In this case,  $\widetilde \phi^{Y}(k)$ has the
form
\begin{equation}
\widetilde\phi^{Y}(x)=\frac{4\pi K\sigma^{3}}{x^{3}(z^{2}+x^{2})}\bar{f}(x),
 \label{Yukawa}
\end{equation}
where
\begin{equation}
 \bar{f}(x)=[z^{2}+x^{2}(1+z)]\sin(x)-xz^{2}\cos(x),
 \label{f_x}
\end{equation}
and $x=k\sigma$. Due to symmetry in the YRPM, the Hamiltonian in (\ref{Xi_full_1}) does not include direct pair interactions
of number density fluctuations.

$\Xi_{\rm{HS}}$ is the grand partition function  of the one-component hard-sphere model with
the dimensional chemical potential $\nu_{\text{HS}}$.
$\Delta\nu_{N}=\bar\nu-\nu_{\text{HS}}$
where
\begin{eqnarray}
\bar\nu=\bar \nu_{\alpha}=\nu_{\alpha}+\frac{\beta}{2V}\sum_{{\mathbf
k}}\widetilde\phi^{Y}(k), \qquad
\alpha=(1,2).
\label{2.7}
\end{eqnarray}
Hereafter, the subscript $\text{HS}$ refers to the hard-sphere system.

The cumulants ${\mathfrak{M}}_{n}^{(i_{n})}$ are expressed in terms of the Fourier
transforms of the  connected correlation functions of the
hard-sphere system \cite{Pat-Mryg-CM}.
$\delta_{{\bf{k}}_{1}+\ldots +{\bf{k}}_{n}}$ is the Kronecker symbol.
In the case of the YRPM, we have the following recurrence relations for the cumulants ${\mathfrak{M}}_{n}^{(i_{n})}$
\cite{Pat-Mryg-CM}:
\begin{eqnarray*}
{\mathfrak{M}}_{n}^{(0)}&=&{\widetilde G}_{n,{\text HS}}, \qquad
{\mathfrak{M}}_{n}^{(1)}=0, \nonumber \\
{\mathfrak{M}}_{n}^{(2)}&=& {\widetilde G}_{n-1,{\text HS}},
\qquad
{\mathfrak{M}}_{n}^{(3)}=0, \nonumber \\
{\mathfrak{M}}_{n}^{(4)}&=&3{\widetilde G}_{n-2,{\text HS}}-2{\widetilde G}_{n-1,{\text HS}},
\end{eqnarray*}
where  ${\widetilde G}_{n,{\text
HS}}$ denotes   the Fourier
transform of the $n$-particle connected correlation function of a
one-component hard-sphere system.
In general,  the dependence of ${\widetilde G}_{n,{\text
HS}}$ on the wave numbers $k_{i}$ is very complicated. Hereafter we use the
following approximation  for ${\widetilde G}_{n,{\text HS}}$
\begin{eqnarray*}
&{\widetilde G}_{2,{\text HS}}(k)\simeq  {\widetilde G}_{2,{\text HS}}(0)+\displaystyle\frac{k^{2}}{2}
{\widetilde G}_{2,{\text HS}}^{(2)}, \\ \nonumber
&{\widetilde G}_{n,{\text HS}}(k_{1},\ldots,k_{n})\simeq
{\widetilde G}_{n,{\text HS}}(0,\ldots) \quad {\text{for}} \quad
n\geq 3, \nonumber
\end{eqnarray*}
where the superscript $(2)$  denotes the second-order derivative with respect to
the wave vector.

 \subsection{Gaussian approximation}
Now we consider the Gaussian approximation of $\Xi$ setting in
Eq.~(\ref{Xi_full_1}) ${\mathfrak{M}}_{n}^{(i_{n})}\equiv 0$ for
$n\geq 3$. Then, after integration over $\omega_{{\bf{k}},N}$  and
$\omega_{{\bf{k}},Q}$ we obtain
\begin{eqnarray*}
\Xi_{{\text G}}&=&\Xi'\int ({\rm d}\rho)
\exp\Big\{\Delta\nu_{N}\rho_{0,N}-\frac{1}{2}\sum_{\bf
k}\left[a_{2}^{(0)}(k)\rho_{{\bf k},N}\rho_{-{\bf k},N} +
a_{2}^{(2)}(k)\rho_{{\bf k},Q}\rho_{-{\bf k},Q}\right]\Big\},
\end{eqnarray*}
where
\begin{displaymath}
\Xi'=\Xi_{\rm{HS}}\exp\left[\Delta\nu_{N}\langle
N\rangle_{\rm{HS}}\right]\prod_{\mathbf
k}\left[{\mathfrak{M}}_{2}^{(0)}{\mathfrak{M}}_{2}^{(2)}\right]^{-1/2},
\end{displaymath}
and
\begin{equation}
a_{2}^{(0)}(k)=\left[{\mathfrak{M}}_{2}^{(0)}(k)\right]^{-1},
\qquad a_{2}^{(2)}(k)=\frac{\beta}{V}\widetilde \phi^{Y}(k)+\left[{\mathfrak{M}}_{1}^{(0)}\right]^{-1}.
\label{C_gaus}
\end{equation}

It follows from Eq.~(\ref{C_gaus}) that $a_{2}^{(0)}(k)$  never vanishes for physical values of the density.
The fact that the YRPM  like the RPM does not undergo the gas-liquid instability in the Gaussian approximation
is due to the absence of direct pair interactions  of density fluctuations  as well as to the neglect of the effect  of non-direct
correlations via a charge subsystem at this level of consideration.
By contrast,  $a_{2}^{(2)}(k)$ can be equal to  zero  at $k=k^{*}\neq 0$, where $k^{*}$ is determined from the condition
$\partial a_{2}^{(2)}/\partial k=0$. The locus  in the phase diagram at which  $a_{2}^{(2)}(k=k^{*})=0$
is called the $\lambda$-line \cite{ciach1,patsaha_mryglod-04} in order to distinguish it from the spinodal line for which $k^{*}=0$.
On the $\lambda$-line the fluid  becomes unstable with respect to the charge ordering indicating that
there can be a phase transition to an ordered phase.
For the RPM ($z=0$), it was found  that  in the presence of fluctuations
the $\lambda$-line disappears and, instead, a first-order phase transition to  an ionic crystal appears \cite{ciach-patsahan:1}.

\subsection{Effective Ginzburg-Landau Hamiltonian }

We consider the model (\ref{2.1}) near the gas-liquid critical
point. In this case,   the phase space of CVs $\rho_{{\bf k},N}$
contains CV $\rho_{0,N}$ related to the order parameter. In order
to obtain the effective Hamiltonian in terms of  $\rho_{{\bf k},N}$,
one should integrate in Eq.~(\ref{Xi_full_1}) over CVs   $\omega_{{\bf k},N}$, $\omega_{{\bf
k},Q}$, and $\rho_{{\bf k},Q}$. A detailed derivation of this type of  Hamiltonian is
given in Ref.~\cite{Patsahan:13}.  Using the results of
Ref.~\cite{Patsahan:13}, we can write an expression for the
effective $\varphi^{4}$ LG Hamiltonian of the model under
consideration
\begin{eqnarray}
&&{\cal H}^{eff}=a_{1,0}\rho_{0,N}+\frac{1}{2!\langle
N\rangle}\sum_{{\mathbf{k}}}\left(a_{2,0}+k^{2}a_{2,2}\right)\rho_{{\bf
k},N}\rho_{-{\bf k},N}+\frac{1}{3!\langle
N\rangle^{2}}\sum_{{\mathbf{k}}_{1},{\mathbf{k}}_{2}} a_{3,0}
\nonumber \\
&& \times\rho_{{\bf k_{1}},N}\rho_{{\bf k_{2}},N}\rho_{-{\bf
k_{1}}-{\bf k_{2}},N}+\frac{1}{4!\langle
N\rangle^{3}}\sum_{{\mathbf{k}}_{1},{\mathbf{k}}_{2},{\mathbf{k}}_{3}}a_{4,0}\rho_{{\bf
k_{1}},N}\rho_{{\bf k_{2}},N}\rho_{{\bf k_{3}},N}\rho_{-{\bf
k_{1}}-{\bf k_{2}}-{\bf k_{3}},N}
\label{H_eff}
\end{eqnarray}
with  the coefficients having the following form in a one-loop
approximation:
\begin{eqnarray}
a_{1,0}&=&-\Delta\nu_{N}-\widetilde{\cal C}_{1,\text{Y}}
\label{a10}\\
a_{n,0}&=&-\rho^{n-1}\,\widetilde {\cal
C}_{n,\text{HS}}-\rho^{n-1}\,\widetilde {\cal
C}_{n,\text{Y}}
\label{an0} \\
a_{2,2}&=&-\frac{1}{2}\rho\,\widetilde {\cal
C}_{2,\text{HS}}^{(2)}-\frac{1}{4\langle
N\rangle}\sum_{\mathbf{q}}\widetilde g_{Y}^{(2)}(q)\left[1+\widetilde
g_{Y}(q)\right].
\label{a22}
\end{eqnarray}
Here, we introduce the following notations. The superscript $(2)$ in
Eq.~(\ref{a22}) denotes the second-order derivative with respect to
the wave vector.  $\widetilde{\cal C}_{n,\text{HS}}$ is the Fourier
transform of the $n$-particle direct correlation function of a
one-component hard-sphere system at $k=0$, and $\rho=\langle N\rangle/V$ is the number density.
Explicit expressions for   $\widetilde{\cal C}_{n,\text{HS}}$ and
$\widetilde {\cal C}_{2,\text{HS}}^{(2)}$ for $n\leq 4$ in the
Percus Yevick (PY) approximation are given in
Ref.~\cite{Patsahan:13}  (see Appendix in Ref.~\cite{Patsahan:13}).

The second term on the right-hand side of Eqs.~(\ref{a10})--\ref{a22}) arises from  the integration over
 CVs  $\rho_{{\bf k},Q}$ and $\omega_{{\bf k},Q}$.
In particular,   $\rho^{n-1}\widetilde{\cal C}_{n,\text{Y}}$ reads
\begin{eqnarray}
\rho^{n-1}\widetilde{\cal C}_{n,
\text{Y}}&=&\frac{(n-1)!}{2}\frac{1}{\langle
N\rangle}\sum_{\mathbf{q}}\left[\widetilde g_{Y}(q)\right]^{n},
\label{Cn_C}
\end{eqnarray}
where
\begin{eqnarray}
\widetilde g_{Y}(q)&=&-\frac{\beta\rho \widetilde\phi^{Y}(q)}{1+\beta\rho
\widetilde\phi^{Y}(q)}
\label{g_q}
\end{eqnarray}
with $\widetilde\phi^{Y}(q)$  given by Eq.~(\ref{Yukawa}).

Taking into account Eqs.~(\ref{Yukawa}) and (\ref{g_q}), one can  obtain
the following explicit expressions for $\rho^{n-1}\widetilde{\cal
C}_{n,\text{Y}}$:
\begin{equation}
-\rho^{n-1}\widetilde{\cal C}_{n,\text{Y}}=\frac{(n-1)!(-24\eta)^{n-1}}{\pi}\int_{0}^{\infty}\,x^{2}
\left[\frac{\bar{f}(x)}{T^{*}x^{3}(z^{2}+x^{2})
+24\eta\bar{f}(x)}\right]^{n}{\rm  d}x,
\label{in}
\end{equation}
where $\bar{f}(x)$ is given by Eq.~(\ref{f_x}).
Hereafter,   the following reduced units are introduced for the
temperature
\begin{equation}
 T^{*}=(\beta K)^{-1}
\label{cr_temp}
 \end{equation}
 and for the density
 \begin{equation}
 \eta=\frac{\pi}{6}\rho^{*}, \quad
 \rho^{*}=\rho\sigma^{3}.
 \label{cr_dens}
\end{equation}

The explicit expression  for the second term in Eq.~(\ref{a22}) is too long to be presented herein. We only emphasize  that
although the Hamiltonian in Eq.~(\ref{Xi_full_1}) does not include direct  pair interactions of number density fluctuations, the effective
short-range attraction does appear in the effective Hamiltonian (\ref{H_eff}). Moreover, in
the limit of charged point particles, i.e., $z=0$ and  $\sigma=0$,  the expression for $a_{2,2}$ leads to the correct result for
the density-density correlation length (see Refs.~\cite{Patsahan:13,Lee_Fisher:96}).

The term $\Delta\nu_{N}$ in Eq.~(\ref{a10})  can be rewritten
as follows [see Eq.~(\ref{2.7})]:
\begin{equation}
\Delta\nu_{N}=\nu-\nu_{\text{HS}}+\frac{1}{2T^{*}}. \label{delta_nu}
\end{equation}

Summarizing, the expressions for coefficients $a_{2,0}$, $a_{3,0}$,
$a_{4,0}$, and $a_{2,2}$ consist of two terms. While the first term
depends solely on the characteristics of a hard-sphere system,  the
second term  is of a mixed type and takes into account the charge-charge
(concentration-concentration) correlations.
Coefficient $a_{1,0}$ is the excess part of the chemical potential
$\nu$, and the equation $a_{1,0}=0$ yields  the chemical potential in
a one-loop approximation. It follows from Eqs.~(\ref{a10}), (\ref{in}) and
(\ref{delta_nu}) that
\begin{equation}
\nu= \nu_{\text{HS}}-\frac{1}{2T^{*}}+\frac{1}{\pi}\int_{0}^{\infty}\frac{x^{2}{\bar f}(x)}{T^{*}x^{3}(z^{2}+x^{2})
+24\eta\bar{f}(x)}{\rm  d}x.
\label{nu_rpa}
\end{equation}
where $\nu_{\text{HS}}$ includes   ideal and  hard-sphere parts. Using the above equation, one can obtain the gas-liquid diagram in
the mean-field approximation.

\section{Gas-liquid phase transition}

In this section we study the gas-liquid phase diagram of the model (\ref{2.1}) using  Eq.~(\ref{an0}) and
Eqs.~(\ref{in})-(\ref{nu_rpa}).
First, we consider the critical point. At the critical point, the  system of equations
 \begin{eqnarray}
  a_{2,0}(\rho_{c},T_{c})=0, \qquad a_{3,0}(\rho_{c},T_{c})  =0
\label{cr-point}
 \end{eqnarray}
holds  yielding the critical temperature and the critical density for the fixed value of $z$. Using
Eqs.~(\ref{f_x}) and (\ref{in}), these equations  can be rewritten as follows:
\begin{eqnarray}
\frac{(1+2\eta)^{2}}{(1-\eta)^{4}}&-&\frac{24\eta}{\pi}\int_{0}^{\infty}\frac{x^{2}{\bar f}^{2}(x){\rm  d}x}{[T^{*}x^{3}(z^{2}+x^{2})
+24\eta\bar{f}(x)]^{2}}=0, \label{a2_zero} \\
\frac{(1-7\eta-6\eta^{2})(1+2\eta)}{(1-\eta)^{5}}&-&\frac{1152\eta^{2}}{\pi}\int_{0}^{\infty}\frac{x^{2}{\bar f}^{3}(x){\rm  d}x}
{[T^{*}x^{3}(z^{2}+x^{2})
+24\eta\bar{f}(x)]^{3}}=0. \label{a3_zero}
\end{eqnarray}
Here, the PY approximation   is used for  $\widetilde{\cal C}_{n,\text{HS}}$. It is worth noting that
Eq.~(\ref{a2_zero}) yields the spinodal curve.

Solving Eqs.~(\ref{a2_zero}) and(\ref{a3_zero})  we obtain the critical
temperature $T_{c}^{*}$ and the critical density $\rho_{c}^{*}$ for  $z$ ranging from $z=0.001$
to $z=2.781$. At $z\geq 2.782$, the system of equations
(\ref{a2_zero}) and (\ref{a3_zero}) has no solution in the region of the
gas-liquid phase transition indicating a disappearance of the critical point.
The dependence of $T_{c}^{*}$ and
$\rho_{c}^{*}$ on the parameter $z^{-1}$ measuring the interaction
range is displayed in Figs.~2 and 3, respectively. As is seen, the
reduced critical temperature $T_{c}^{*}$ rapidly decreases  with an
increase of the interaction range for $z^{-1}\leq 20$ and then
slowly approaches   the critical temperature of the RPM
($T_{c}^{*}=0.08446$). The reduced critical density $\rho_{c}^{*}$
demonstrates a sharp decrease in the region $z^{-1}\leq 10$ reaching
the RPM critical value for $z^{-1}\simeq 100$. A decrease of both the critical
temperature and the critical density expressed in the same reduced
units is observed in Ref.~\cite{Fortini:06}.

 \begin{figure}[htbp]
 \centering
 \includegraphics[height=6cm]{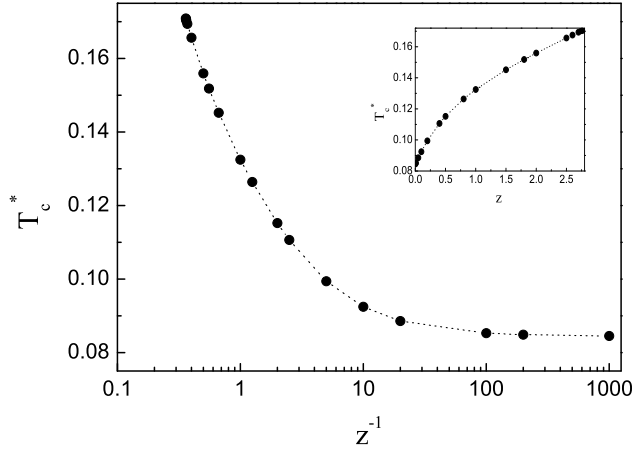}
 \caption{Reduced critical temperature $T_{c}^{*}$ [Eq.~(\ref{cr_temp})] of the YRPM as a function of the interaction range.
 The inset shows
  $T_{c}^{*}$ as a function of $z$. The line is a guide to the eye.} \label{fig2}
 \end{figure}
 \begin{figure}[htbp]
 \centering
 \includegraphics[height=6cm]{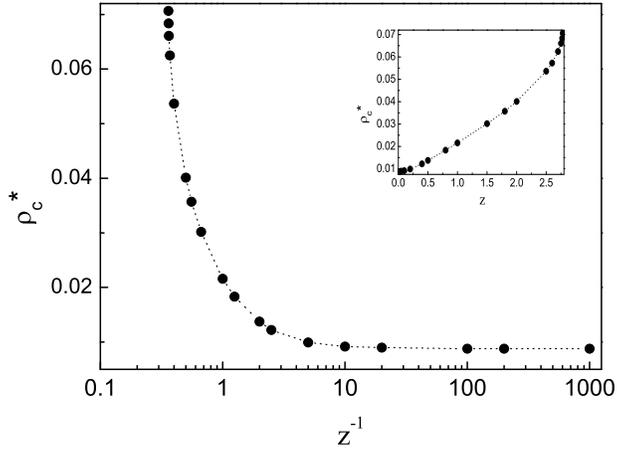}
 \caption{Reduced critical density [Eq.~(\ref{cr_dens})] of the YRPM as a function of the interaction range. The inset shows
  $\rho_{c}^{*}$ as a function of $z$. The line is a guide to the eye.
 } \label{fig3}
 \end{figure}

 We calculate the  spinodal curves for
different values of $z$ using Eq.~(\ref{a2_zero}).  The results are
presented in the ($T^{*}$,$\eta$) plane  in Fig.~4. As is seen, the
spinodals change their shape with the variation of the interaction
range. For  small values of $z$, the curves have a noticeable
maximum at small $\eta$ and change their run passing through a
minimum. The maximum point of the spinodal coincides with the gas-liquid
critical point. The second positive slope of spinodal curves
appearing at higher densities indicates another type of  phase
instability
induced by  the charge ordering.
We suggest that this branch of  the spinodal   should be an indication of the pretransitional effects associated
with  crystallization.
For the system of oppositely charged colloids, a
broad fluid--${\rm CsCl}$ crystal phase coexistence is found
experimentally  \cite{Hynninen-06} and by computer simulations
\cite{Hynninen-06,Fortini:06}.   Moreover, it is shown that
fluid-solid phase diagrams of the YRPM  and the RPM are
qualitatively  similar \cite{Hynninen-06}. When $z$ increases, the
maximum of  spinodals moves to  higher densities, becomes flatter
and finally disappears at $z>2.781$. At  $z=2.781$, the gas-liquid
critical point merges with the spinodal branch  induced by the
charge ordering.

 \begin{figure}[htbp]
 \centering
 \includegraphics[height=6cm]{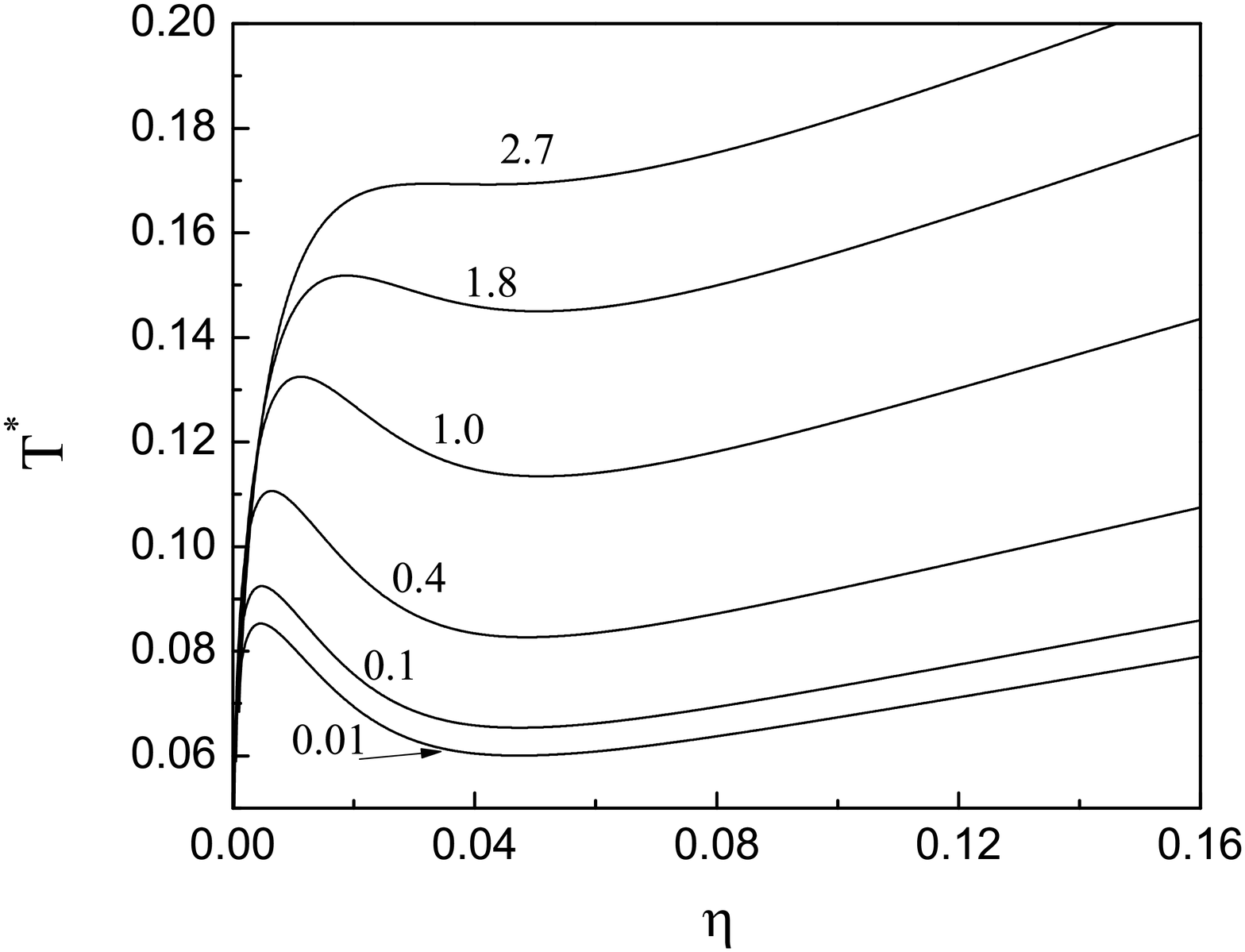}
 \caption{Spinodal curves  of the YRPM for  $z$ ranging from $0.01$ to $2.7$
 in the ($T^{*}$,$\eta$) representation.
 } \label{fig4}
 \end{figure}
 \begin{figure}[htbp]
 \centering
 \includegraphics[height=6cm]{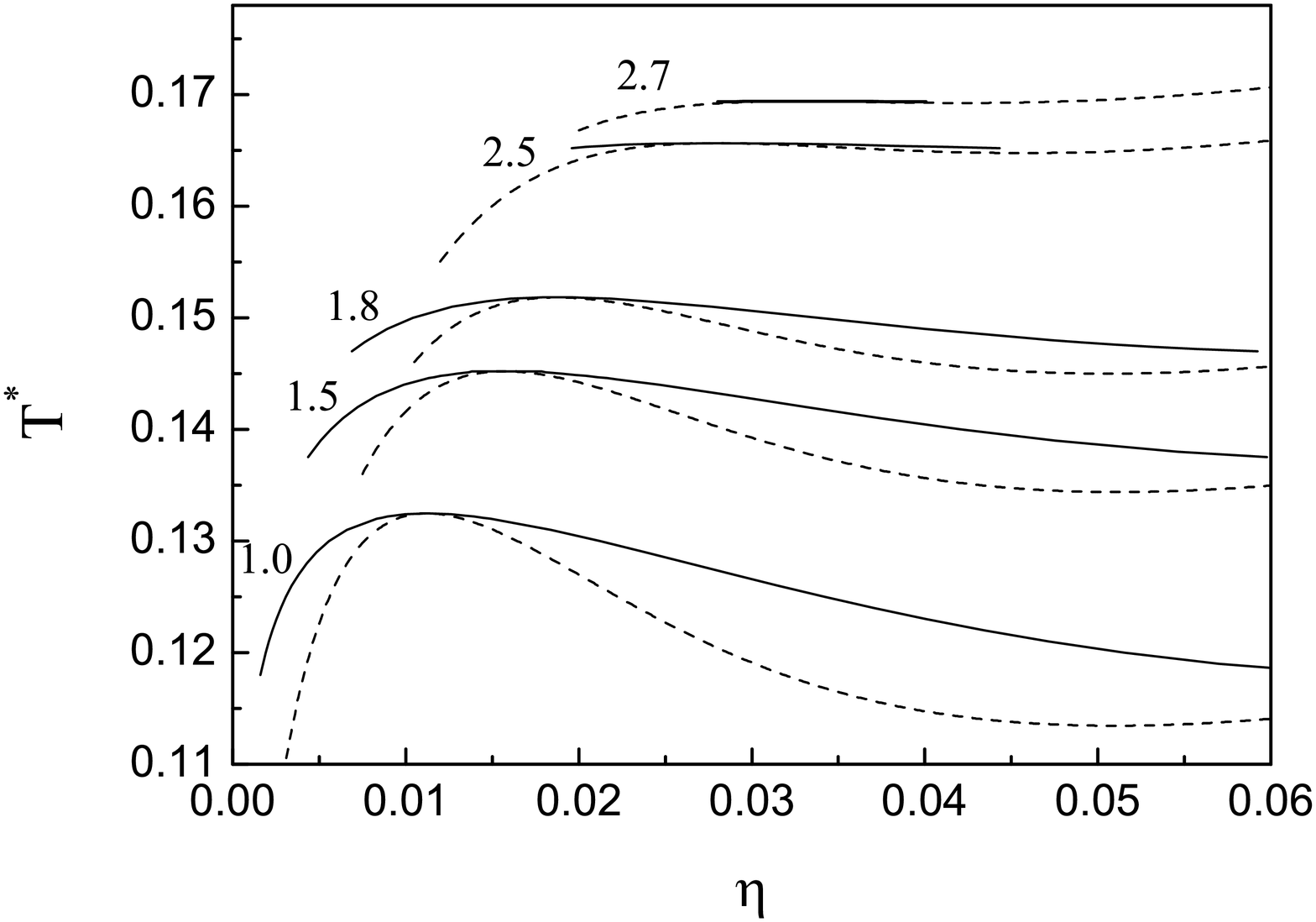}
 \caption{Coexistence  curves (solid) and spinodal curves (dashed) of the YRPM for  $z$
 ranging from $1.0$ to $2.7$ in the ($T^{*}$,$\eta$)
 representation.
 } \label{fig5}
 \end{figure}

To calculate the  coexistence curves, we use Eq.~(\ref{nu_rpa}) for the chemical potential and  employ the Maxwell
 double-tangent construction. Figure~5 shows  both the  coexistence curves (solid lines) and  spinodals (dashed lines)
 in the ($T^{*}$,$\eta$) plane
 for a set of  $z$ values.  As is seen,
 the region of  gas-liquid coexistence   reduces  with an increase of
 $z$. Furthermore, the coexistence curves become very flat for $z\geq 2.7$. This means that  the liquid phase  becomes
 more and more difficult to observe in this domain of $z$.
 For $z>2.781$, no critical point  can be calculated and $z=2.781$ can be considered as the limit value for  gas-liquid
 phase separation in the  approximation considered in this paper. We recall that the limit value for a stable  gas-liquid
 separation  obtained in simulations
 is $z=4$ \cite{Fortini:06}.

 \section{The crossover temperature}

In this section,  we study the effect of the interaction range on
the temperature region in which the crossover from classical
behavior to Ising-like critical behavior occurs.  To this end, we
use the Ginzburg criterion  \cite{levanyuk,ginzburg}. This criterion
defines the reduced Ginzburg temperature $t_{G}$ which marks a lower
bound of the temperature region where   a mean-field description is
self-consistent. For    $|t|\ll t_{G}$  where $|t|=|T-T_{c}|/T_{c}$,
Ising critical behavior should be exhibited. Therefore, it is
reasonable to take the reduced Ginzburg temperature as an estimate
of the crossover temperature \cite{Gutkowskii-Anisimov,fisher3,Chaikin_Lubensky}.

The Ginzburg temperature expressed in terms of  coefficients of the Hamiltonian (\ref{H_eff}) reads~\cite{fisher3}
\begin{eqnarray}
 t_{G}=\displaystyle\frac{1}{32\pi^{2}}\frac{a_{4,0}^{2}}{a_{2,t}
 a_{2,2}^{3}},
\label{t_G}
\end{eqnarray}
where $a_{2,t}=\left.\partial a_{2,0}/\partial t\right|_{t=0}$. Taking into account Eqs.~(\ref{an0}) and (\ref{in}), one can
obtain for $a_{2,t}$
 \begin{eqnarray}
 a_{2,t}=\frac{48\eta T_{c}^{*}}{\pi}\,\int_{0}^{\infty}\frac{x^{5}(z^{2}+x^{2}){\bar{f}}^{2}(x)}{\left(T_{c}^{*}x^{3}(z^{2}+x^{2})
 +24\eta\bar{f}(x)\right)^{3}}{\rm
 d}x,
 \label{a2t_yrpm}
 \end{eqnarray}
where $\bar{f}$ is given by (\ref{f_x}).

 \begin{figure}[h]
 \centering
 \includegraphics[height=6cm]{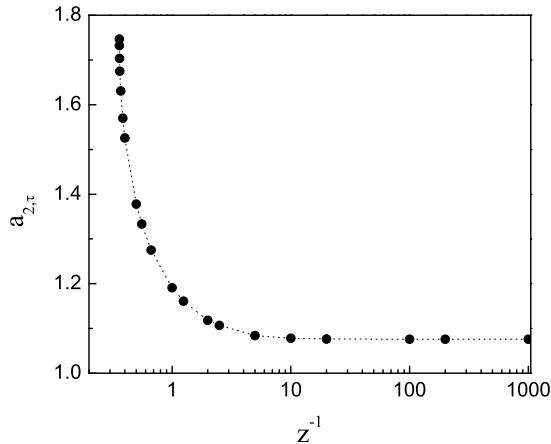}
 \caption{The coefficient $a_{2,t}$ as a function of the interaction range
 $z^{-1}$. The line is a guide to the eye.
  } \label{fig6}
 \end{figure}

The relevant coefficients of the LG Hamiltonian are  calculated at $T^{*}=T_{c}^{*}$ and $\rho^{*}=\rho_{c}^{*}$ using
Eqs.~(\ref{an0}), (\ref{a22}), (\ref{in}), and
(\ref{a2t_yrpm}).
It is instructive to view the coefficients $a_{2,t}$, $a_{2,2}$, and $a_{4,0}$ as functions of $z^{-1}$.
Figures~6--8 show the dependence  of   coefficients
on the interaction range. While $a_{2,t}$ is a
decreasing function of $z^{-1}$, the  other two coefficients
demonstrate a nonmonotonous  behavior.  It is worth noting that $a_{2,t}>1$ for the whole range of $z$ for which
coexistence exists. The coefficient $a_{2,2}$ corresponds to a squared range
of the effective density-density attraction.
Being nearly constant  for
$z\leq 0.1$, $a_{2,2}$   decreases for larger values of $z$
and attains  a minimum at $z\simeq 1.8$.  Then, it slightly increases  in the range $1.8< z<2.78$.  The coefficient $a_{4,0}$
has a maximum at $z\simeq 1.5$ and then (for $z>1.5$) sharply tends
to zero   indicating the
presence of a tricritical point  at $z=2.781$ for which our estimate is $T_{c}^{*}=0.1709$,
$\rho_{c}^{*}=0.0718$.
For $z\lesssim 0.01$ ($z^{-1}\gtrsim 100$),
all three coefficients become equal to the
corresponding coefficients of the RPM \cite{Patsahan:13}.

 \begin{figure}[h]
 \centering
 \includegraphics[height=6cm]{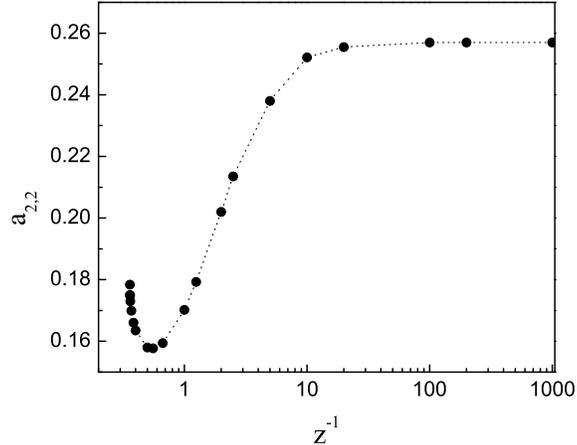}
 \caption{Coefficient $a_{2,2}$ as a function of the interaction range  $z^{-1}$ The line is a guide to the eye.
 } \label{fig7}
 \end{figure}

 \begin{figure}[h]
 \centering
 \includegraphics[height=6cm]{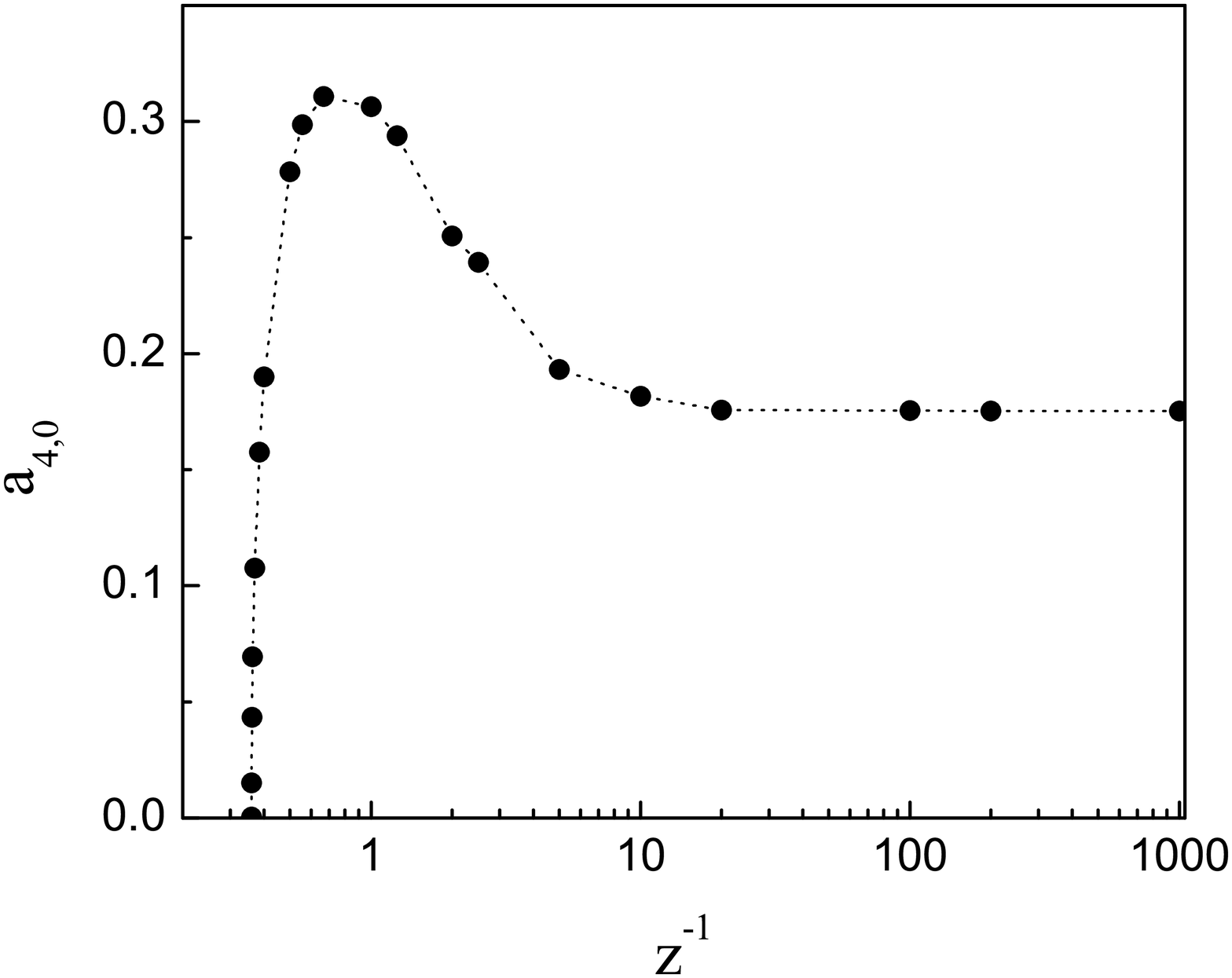}
 \caption{Coefficient $a_{4,0}$ as a function of the interaction range
 $z^{-1}$. The line is a guide to the eye.
 } \label{fig8}
 \end{figure}

 \begin{figure}[h]
 \centering
 \includegraphics[height=7cm]{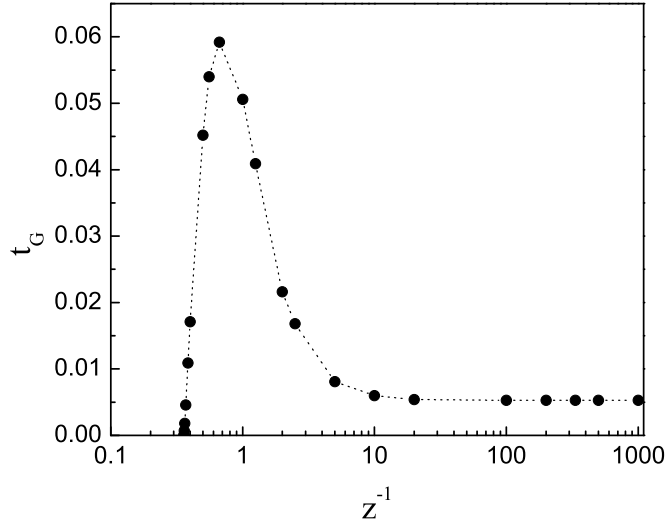}
 \caption{Reduced Ginzburg temperature as a function of the interaction range
 $z^{-1}$. The line is a guide to the eye.
 } \label{fig9}
 \end{figure}

The dependence  of the reduced Ginzburg temperature $t_{G}$ on the
interaction range is shown in Fig.~9. For $z\simeq 0.01$ ($z^{-1}\simeq 100$), the
reduced Ginzburg temperature approaches the value $t_{G}=0.0053$ obtained for the
RPM \cite{Patsahan:13}. For large values of $z$ (small $z^{-1}$), $t_{G}$ shows a nonmonotonous behavior
passing through a sharp maximum at $z\simeq 1.5$ and approaching
zero at $z\simeq 2.78$. Remarkably, a maximum value of $t_{G}$ is about $10$ times larger  than that obtained for the RPM.

\section{Conclusions}

Using the approach that exploits the method of CVs we have studied
the gas-liquid coexistence and the associated  crossover behavior in the
screened Coulomb  restricted primitive model (YRPM). For this model, we have
obtained explicit expressions for all the relevant coefficients of
the LG Hamiltonian in a one-loop approximation.  Gas-liquid phase diagram,  critical parameters
and  Ginzburg temperature  are calculated for $0.001\leq z\leq
2.781$ using these
expressions. It  should be emphasized that the approximation considered produces
the mean-field phase diagram.

First, we have studied the dependence of  critical temperature and critical
density  on the interaction range of the Yukawa potential. The critical temperature scaled by the Yukawa potential
contact value increases  with an increase
of  the inverse screening length  for the whole range of $z$ for which coexistence exists.
The reduced critical density shows a
similar trend. Both trends   qualitatively agree with the results
of simulations \cite{Fortini:06}. A rapid increase in the critical temperature and  density above
the corresponding values of the RPM (up to $z\approx 4$)
was also found theoretically using the MSA and the GMSA \cite{Carvalho_Evans:97,Mier-Y-Teran}.

As for the gas-liquid phase diagram, our results have shown that the region of  coexistence
in the  temperature-density plane  reduces with an
increase of the inverse screening length $z$  and completely
disappears  at $z> 2.78$.
The trend of the evolution of   gas-liquid coexistence
with the variation of $z$ is generally   consistent with the
results of computer simulations  indicating a stable gas-liquid separation  for $z\leq 4$ \cite{Fortini:06}.
However,   the gas-liquid binodal obtained in simulations does not disappear  but becomes
metastable with respect to the solid-fluid separation for $z>4$.
In this study, we have  focused exclusively
on the gas-liquid equilibrium. The description of transitions involving a solid phase requires going beyond
the treatment we have presented here. This issue will be addressed elsewhere.

Finally, we have studied the effect of the interaction region on the
crossover behavior by applying the Ginzburg criterion. We have
analyzed the coefficients of the LG Hamiltonian as functions of the
interaction range. It is significant that for $z\leq 0.01$, all the
coefficients approach the values obtained for the RPM. It appears
that the coefficient $a_{4,0}$ decreases for $z>1.5$ and approaches
zero when $z\simeq 2.78$ indicating the existence of a tricritical
point. Accordingly, the reduced Ginzburg temperature  tends to zero
in this domain of $z$.  In this case, the tricritical point is the point where the
gas-liquid critical point merges with the spinodal branch induced
by the charge ordering. The possible existence  of a tricritical
point for the YRPM with a large $z$ was discussed in
Ref.~\cite{stell1}. For $z<2.78$, $t_{G}$ shows a nonmonotonous
behavior. First, $t_{G}$ increases reaching a maximum at $z\simeq
1.5$ and then for $z< 1.5$, $t_{G}$ again decreases approaching the
RPM value for $z\simeq 0.01$.  It is interesting to  note that the
reduced Ginzburg temperature for the YRPM with $z=1.8$ is about $10$
times larger than $t_{G}$ for the RPM ($z=0$). Therefore, we have
found that an increase in the interaction region from the one
typical of nonionic fluids to the one typical of ionic fluids  leads
to a decrease of the temperature region where the crossover from the
mean-field critical behavior to Ising model criticality occurs.
Extending our previous studies, we have clearly demonstrated that
the range of the interactions plays a crucial role in the crossover
behavior observed in ionic fluids.



%
\end{document}